\documentclass[prd,a4paper,preprint,preprintnumbers,nofootinbib]{revtex4-2}

\usepackage[a4paper, hdivide={1.91cm,,1.165cm}, vdivide={1.83cm,,3.0cm}]{geometry}

\usepackage{amsfonts}
\usepackage{braket}
\usepackage{physics}
\usepackage{amsmath}
\usepackage{graphicx}
\usepackage{xspace}
\usepackage{color}
\usepackage{units}
\usepackage{slashed}
\usepackage[hyperfootnotes=false]{hyperref}
\hypersetup{linkcolor=red,
citecolor=green,
filecolor=cyan,
urlcolor=magenta}
\usepackage{gensymb}
\usepackage{booktabs}
\usepackage{multirow}
\usepackage{xcolor}
\usepackage{setspace}
\allowdisplaybreaks

\begin{document}


\title{Charmed baryon $\Omega_c^0 \rightarrow \Omega^- l^+ \nu_l$ and $\Omega_c^0 \rightarrow \Omega_c^- \pi^+ (\rho^+)$ decays in light cone sum rules}

\author{T.~M.~Aliev}
\email{taliev@metu.edu.tr}
\affiliation{Department of Physics, Middle East Technical University, Ankara, 06800, Turkey}

\author{S.~Bilmis}
\email{sbilmis@metu.edu.tr}
\affiliation{Department of Physics, Middle East Technical University, Ankara, 06800, Turkey}
\affiliation{TUBITAK ULAKBIM, Ankara, 06510, Turkey}

\author{M.~Savci}
\email{savci@metu.edu.tr}
\affiliation{Department of Physics, Middle East Technical University, Ankara, 06800, Turkey}

\date{\today}

\begin{abstract}
The semileptonic $\Omega_c^0 \rightarrow \Omega^- l \nu$ and non-leptonic $\Omega_c^0 \to \Omega^- \pi^+$, $\Omega_c^0 \to \Omega^- \rho^+$ decays of charmed $\Omega_c$ baryon are studied within the light cone sum rules. The form factors responsible for $\Omega_c \to \Omega$ transitions are calculated using the distribution amplitudes of $\Omega_c$ baryon. With the obtained form factors, the branching ratios of $\Omega_c^0 \to \Omega^- l^+ \nu_l$, $\Omega_c^0 \rightarrow \Omega^- \pi^+$, and $\Omega_c^0 \rightarrow \Omega^- \rho^+$ decays are estimated. The results are compared with Belle data as well as the findings of the other approaches.
\end{abstract}

\maketitle

\newpage

%
\section{Introduction\label{intro}}
\label{sec:1}
The semileptonic weak decays of hadrons represent a very promising class of decays. The study of semileptonic decays can provide us with useful information about the elements of the Cabibbo - Kobayashi- Maskawa (CKM) matrix. The investigation of these decays can play a crucial role in studying strong interaction, i.e., the form of the effective Hamiltonian.
The decay amplitudes of semileptonic decays can be represented as a product of a
well-understood leptonic current and a complicated hadronic current for describing
the quark transitions. The hadronic part of the weak decays is usually  parameterized
in terms of form factors. The form factors belong to the nonperturbative region of
QCD, hence some nonperturbative methods are needed to calculate them. Among these methods, the QCD sum
rules~\cite{Shifman:1978by} occupies a special place. The advantage of this
method is that it is based on the fundamental QCD Lagrangian.

The lowest lying $\Omega_c$ baryon predominantly decays weakly~\cite{ParticleDataGroup:2020ssz}. Up to now, several $\Omega_c^0$ decays, such as $\Omega_c^0 \rightarrow \Xi^0 \bar{K}^{(*)0}$, $\Omega^- \rho^+$ and $\Omega^- l^+ \nu_e$ decays are observed~\cite{CLEO:2002imi}. First observation of semileptonic decays $\Omega_c \rightarrow \Omega^- e^+ \nu_e$ was achieved by CLEO collaboration~\cite{CLEO:2002imi} with $ \mathcal{R} = \frac{\mathcal{B} (\Omega_c^0 \rightarrow \Omega^- e^+ \nu_e)}{ \mathcal{B}(\Omega_c^0 \rightarrow \Omega^- \pi^+)} = 2.4 \pm 1.2$. This decay has been carefully investigated within different approaches such as the light-front quark model~\cite{Hsiao:2020gtc,Huang:2021ots}, heavy quark expansion model~\cite{Voloshin:1996vb} and quark model~\cite{Pervin:2006ie}. However, the predictions of the branching ratios of $\Omega_c^0 \rightarrow \Omega^- l^+ \nu_e$ varies between $0.005$ and $0.127$, and this large variation deserves much more attention.

Recently, Belle Collaboration has announced the first observation of $\Omega_c^0 \to \Omega^- \mu^+ \nu_\mu$ decay~\cite{Belle:2021dgc}. In this study, measurement of the branching ratios of $\Omega_c^0 \to \Omega^- l^+ \nu_l $ $(l = e \text{ or } \mu)$ decays are compared to the reference mode $\Omega_c^0 \to \Omega^- \pi^+$, namely the branching ratios $\frac{\mathcal{B}(\Omega_c^0 \to \Omega^- e^+ \nu_e)}{\mathcal{B}(\Omega_c^0 \to \Omega^-  \pi^+)}$ and $\frac{\mathcal{B}(\Omega_c^0 \to \Omega^- \mu^+ \nu_\mu)}{\mathcal{B}(\Omega_c^0 \to \Omega^- \pi^+)}$ to be $1.98 \pm 0.13 (\text{stat)} \pm \text{0.08 (\text{syst})}$ and $1.94 \pm 0.18 (\text{stat}) \pm 0,10 (\text{syst})$, respectively.

The new measurement and the variation in the predictions of the branching fractions among different models need further attention. In the present work, we study the $\Omega_c^0 \to \Omega^- l^+ \nu_e$ decay within the light cone sum rules (LCSR) method (for more information about the LCSR method, see~\cite{Braun:1997kw}).

The paper is organized as follows. In Sec.~\ref{sec:2}, the LCSR for the relevant form factors responsible for $\Omega_c^0 \to \Omega^-$ transitions are derived. Numerical analysis, including the results for the form factors and decay widths, is presented in Sec.~\ref{sec:3} . The last section contains our conclusion.  
\section{Form Factors for $\Omega_c \rightarrow \Omega$ transition in LCSR}
\label{sec:2}
To calculate the form factors for $\Omega_c^0 \to \Omega^-$ transition, we consider the following correlation function
\begin{equation}
  \label{eq:2}
  \Pi_{\mu \nu} = i \int d^4x e^{i p^\prime x} \langle 0 | T \big\{J^{\Omega}_{\mu} J^{V-A}_{\nu}(0) \big\} | \Omega_c  \rangle
\end{equation}
Here, the current $J^\Omega_\mu = \epsilon^{j k l} s^{j^T} C \gamma_\mu s^k s^l$ is the interpolating current of $\Omega$ baryon, $J_\nu^{V-A}(0) = \bar{s} \gamma_\nu (1- \gamma_5) c$ is the current describing $c \to s$ transition and $j, k$ and $l$ are the color indices.

In LCSR, the correlation function is calculated both at hadronic and  QCD level at the deep Euclidean region, i.e., $p^2 << m_c^2$, $q^2 << m_c^2$. Then, the results of the calculations for the two representations of the correlation function are matched by using the quark-hadron duality ansatz. In result, the sum rules for the relevant form factors are derived.

Let us first calculate the correlation function from the hadronic side. Inserting the complete set of baryon states carrying the quantum numbers of $\Omega$ baryon and isolating the ground state for the correlation function, we obtain
\begin{equation}
  \label{eq:3}
  \Pi_{\mu \nu} = \frac{\lambda \langle 0 | J^\Omega_\mu | \Omega(p^\prime) \rangle \langle \Omega(p^\prime) | \bar{s} \gamma_\nu (1-\gamma_5) c| \Omega_c(p) \rangle}{m^2_{\Omega} - p^{\prime^2} }~.
\end{equation}
The first term of this matrix is defined
\begin{equation}
  \label{eq:4}
  \langle 0 | J^\Omega_\mu | \Omega(p^\prime) \rangle = \lambda u_\mu(p^\prime)
\end{equation}
where $\lambda$ is the decay constant and $u_\mu(p^\prime)$ is the Rarita-Schwinger spinor for spin-$3/2$ $\Omega$ baryon.

The second matrix element is parameterized in terms of eight transition form factor
\begin{equation}
  \label{eq:5}
  \begin{split}
    \langle \Omega(p^\prime) | \bar{s} \gamma_\mu (1-\gamma_5) c | \Omega_c(p) \rangle  & = \bar{u}_\alpha(p^\prime) \bigg\{ \bigg[
    \frac{p_\alpha}{m_{\Omega_c}} ( \gamma_\nu F_1 + \frac{p_\nu}{m_{\Omega_c}}F_2 + \frac{p^\prime_\nu}{m_{\Omega}} F_3) + F_4 g_{\nu \alpha} \bigg] \gamma_5 \\
    &- \bigg[ \frac{p_\alpha}{m_{\Omega_c}} ( \gamma_\nu G_1 + \frac{p_\nu}{m_{\Omega_c}}G_2 + \frac{p^\prime_\nu}{m_{\Omega}}) + G_4 g_{\mu \alpha} \bigg]  \bigg\} u(p)~.
\end{split}
\end{equation}
The summation over spins of $\Omega$ baryon is performed via the formula
\begin{equation}
  \label{eq:6}
  \sum_s u_\alpha(p^\prime) \bar{u}_\beta(p^\prime) = - (\slashed{p}^\prime + m_{\Omega_c}) \bigg[ g_{\alpha \beta} - \frac{1}{3} \gamma_\alpha \gamma_\beta - \frac{2}{3} \frac{p_\alpha^\prime p_\beta^\prime}{m^2_{\Omega}} + \frac{1}{3} \frac{p_\alpha^\prime \gamma_\beta - p^\prime_\beta \gamma_\alpha}{m_{\Omega}} \bigg]
\end{equation}
Before delving into the analysis, we would like to make the following two remarks.
\begin{itemize}
\item The current $J^{\Omega}_\mu$ also couples with spin-1/2 negative parity baryons, i.e.,
  \begin{equation}
    \label{eq:11}
    \langle 0 | J^\Omega_\mu | B^- (p^\prime) \rangle \sim \bigg[ \gamma_\mu - \frac{4}{m}p^\prime_\mu \bigg] u(p^\prime,s)
  \end{equation}
  where $B$ denotes the negative parity baryon. Therefore, the structures with $p^\prime_\mu$ and $\gamma_\mu$ terms also contain the contributions of spin-1/2 baryon. Using this fact, from Eq.\eqref{eq:6} we find out that only the structure with $ g_{\alpha \beta}$ term is free of spin-1/2 baryon contributions. In other words, the structure with $ g_{\alpha \beta}$ term contains contributions coming only from spin-3/2 baryons.
\item Note that not all the Lorentz structures are independent, and to overcome this problem, a specific order of Dirac matrices are chosen. In this work, we choose the structure $\gamma_\mu \slashed{q} \gamma_\nu \slashed{p}~(\gamma_\mu \slashed{q} \gamma_\nu \slashed{p} \gamma_5)$.
\end{itemize}
  
  Taking into account these remarks, we obtain the correlation function from the hadronic part,
  \begin{equation}
    \label{eq:12}
    \begin{split}
     \Pi_{\mu \nu} &=  \frac{1}{m_{\Omega}^2 - p^{\prime^2}} \bigg\{ \frac{p_\mu}{m_{\Omega_c}} \big[ F_1 (2p_\nu \gamma_5 + (m_{\Omega_c}+m_\Omega) \gamma_\nu \gamma_5  - \slashed{q} \gamma_\nu  \gamma_5) 
      + F_2 \frac{p_\nu}{m_{\Omega_c}} \big( (m_\Omega-m_{\Omega_c}) \gamma_5 - \slashed{q} \gamma_5 \big) \\
      &+ \frac{1}{m_\Omega}F_3(p-q)_\nu \big( (m_\Omega-m_{\Omega_c}) \gamma_5 - \slashed{q} \gamma_5 \big) \big] 
      + F_4 g_{\mu \nu} \big( (m_\Omega-m_{\Omega_c}) \gamma_5 - \slashed{q}  \gamma_5 \big) \\
      &- \frac{p_\mu}{m_{\Omega_c}} \big[ G_1 \big( 2 p_\nu + (m_\Omega-m_{\Omega_c}) \gamma_\nu - \slashed{q} \gamma_\nu \big) 
      + \frac{G_2 p_\nu}{m_{\Omega_c}} ((m_\Omega + m_{\Omega_c}) - \slashed{q}) \\
      &+ G_3 \frac{(p - q)_\nu}{m_{\Omega}} \big( (m_\Omega + m_{\Omega_c}) - \slashed{q} \big) \big]
      +G_4 g_{\mu \nu} (m_\Omega + m_{\Omega_c}) - \slashed{q}  \bigg\} + ...
    \end{split}
  \end{equation}
where dots stand for the contributions of higher states and continuum, which denotes the contributions arising from quarks starting from some threshold $s_{th}$ value.
In the following discussions, we will denote the momentum of $\Omega_c$ baryon as $p_\mu \to m_{\Omega_c} v_\mu$ where $v_\mu$ is its velocity.

Now let us turn our attention to the calculation of the correlation function from the QCD side with the help of the operator product expansion (OPE). Using the Wick theorem from Eq.~\eqref{eq:2} we get the correlation function
\begin{equation}
  \label{eq:14}
  \begin{split}
    \Pi_{\mu \nu} &= i \int d^4 x e^{i p^\prime x} \langle 0 | \epsilon^{j k l} (C \gamma_\mu)_{\alpha \beta}(\mathcal{T}_\nu)_{\alpha_1 \beta_1} \\
    & \bigg\{ S_{\gamma \alpha_1}^{c j_1}(x) s_\alpha^j(x) s_\beta^k(x) c_{\beta_1}^{j_1}(0) -
    S_{\beta \alpha_1}^{k j_1}(x) s_\alpha^j(x) s_\gamma^l(x) c_{\beta_1}^{j_1}(0) +
    S_{\alpha \alpha_1}^{j j_1}(x) s_\beta^k(x) s_\gamma^l(x) c_{\beta_1}^{j_1}(0) \bigg\} | \Omega_c \rangle 
\end{split}
\end{equation}
where $S(x)$ is the s-quark propagator, and $\mathcal{T}_\nu = \gamma_\nu (1- \gamma_5)$. From Eq.\eqref{eq:14}, it follows that to calculate the correlation function from QCD side, we need the matrix element $\epsilon^{j k l} \langle 0| \bar{s}_{\alpha}^j(x) s_{\beta}^k(x) c_\gamma^l(0) | \Omega_c \rangle$. This matrix element can be parametrized in terms of the heavy baryon distribution amplitudes (DAs). The DAs of the sextet baryons with quantum numbers $J^P = \frac{1}{2}^+$ in the heavy quark mass limit is obtained in~\cite{Ali:2012pn}. In this work, the DA's are classified by the total spin of two light quarks. If the polarization vector is parallel to the light cone plane, the matrix element $\epsilon_{j k l} \langle 0 | q_{1 \alpha}^j (t_1) q_{2 \beta}^k(t_2) Q_\gamma^l(0) | \Omega_Q(p) \rangle $ can be expressed in terms of the four DAs in the following way.
\begin{equation}
  \label{eq:15}
  \begin{split}
    \epsilon_{j k l} \langle 0 | s_{\alpha}^j (t_1) s_{\beta}^k(t_2) c_\gamma^l(0) | \Omega_c(v) \rangle  &= \Sigma A_i (\Gamma_i C^{-1})_{\alpha \beta} (\gamma_5 \bar{\slashed{v}} u_{\Omega_c})_\gamma 
  \end{split}
\end{equation}
where
\begin{align}
  \label{eq:8}
  A_1 &= \frac{1}{8} v_{+} f^{(1)} \psi_2  &    \Gamma_1  &= \bar{\slashed{n}} \nonumber \\
  A_2 &= f^{(2)} \psi_3^{(\sigma)}              &     \Gamma_2  &= \frac{1}{8} i \sigma_{\alpha \beta} \bar{n}_{\alpha} n_{\beta} \nonumber \\
  A_3 &= \frac{1}{4} \psi_3^{(s)} f^{(2)}  &    \Gamma_3  &= 1 \nonumber \\
  A_4 &= - \frac{1}{8v_{+}} \psi_4 f^{(1)} &    \Gamma_4 &= \slashed{n}~.
\end{align}
Here $n_\mu = \frac{x_\mu}{v x}, \bar{n}_\mu = 2 v_\mu - \frac{1}{v x} x_\mu$, $\bar{v}_\mu = \frac{x_\mu}{v x} - v_\mu$, and $f^{(i)}$ are the decay constants of $\Omega_c$ baryon and $\psi^{(i)}$ are the distribution amplitudes. The Fourier transformation of the DAs are $\Psi(x_1,x_2) = \int_0^{\infty} d\omega_1 d\omega_2 e^{-i \omega_1 t_1} e^{-i \omega_2 t_2} \psi(\omega_1 \omega_2)$
where $\omega_1$ and $\omega_2$ are the momentum of two light quarks along the light-cone direction and their total momentum is $\omega = \omega_1 + \omega_2$ with $t_1 = x_1 n$ and $t_2 = x_2 n$.
The DAs can be written as
\begin{equation}
  \label{eq:16}
  \psi(t_1,t_2) = \int_0^{\infty} d \omega \omega \int du e^{-i \omega v x_1} e^{-i \omega \bar{u} (x_2 - x_1)} \psi(\omega,u)~.
\end{equation}
In our case since $x_1 = x_2$, then we get
\begin{equation}
  \label{eq:17a}
  \psi(t,t_2) = \int_0^{\infty} d \omega \omega \int du e^{-i \omega v x} \psi(\omega,u)~.
\end{equation}
Based on the heavy-quark symmetry, we can use the same DAs for the baryon containing charm quark, and b-quark.  In~\cite{Ali:2012pn}, the DAs for $\Omega_b$ baryon are obtained and we used the same DAs for $\Omega_c$ in this work.
\begin{align}
  \label{eq:17}
  \psi_2 (\omega, u)  &= 
\omega^2 u (1 - u) \, \sum_{n = 0}^2
\frac{a_n}{{\epsilon_n}^4} \, 
\frac{C^{3/2}_n (2 u - 1)}{| C^{3/2}_n |^2} \,
{\rm e}^{- \omega/\epsilon_n} , \nonumber \\ 
 \psi_4 (\omega, u) &=
\sum_{n = 0}^2 \frac{a_n}{{\epsilon_n}^2} \, 
\frac{C^{1/2}_n (2 u - 1)}{| C^{1/2}_n |^2} \,
{\rm e}^{- \omega/\epsilon_n} , \nonumber \\
 \psi_3 (\omega, u) &=
\frac{\omega}{2} \, \sum_{n = 0}^2
\frac{a_n}{{\epsilon_n}^3} \, 
\frac{C^{1/2}_n (2 u - 1)}{| C^{1/2}_n |^2} \,
{\rm e}^{- \omega/\epsilon_n} ,
\end{align}
where
\begin{equation}
  \label{eq:20}
\left | C^\lambda_n \right |^2 = 
\int_0^1 du \left [ C^\lambda_n (2 u - 1) \right ]^2 , 
\end{equation}
%
with
$\big | C^{1/2}_0 \big |^2 = \big | C^{3/2}_0 \big |^2 = 1$,
$\big | C^{1/2}_1 \big |^2 = 1/3$, $\big | C^{3/2}_1 \big |^2 = 3$,
$\big | C^{1/2}_2 \big |^2 = 1/5$, and $\big | C^{3/2}_2 \big |^2 = 6$.
The parameters entering to the Eqs.~\eqref{eq:17} are obtained in~\cite{Ali:2012pn} and we present their expressions in Table~\ref{tab:1} for completeness. For the numerical calculations we take $A= 1/2$.
\begin{table*}[htb]
  \centering
  \renewcommand{\arraystretch}{1.4}
  \setlength{\tabcolsep}{7pt}
  \begin{tabular}{ccccccc}
    \toprule
twist & $a_0$ & $a_1$ & $a_2$                     & $\epsilon_0$                & $\epsilon_1$               & $\epsilon_2$                \\
$2$   & $1$   & $-$   & $\frac{8 A+1}{A+1}$       & $\frac{1.3A+1.3}{A+6.9}$    & $-$                        & $\frac{0.41A+0.06}{A+0.11}$ \\
$3 \sigma$ & $1$   & $-$   & $\frac{0.17 A-0.16}{A-2}$ & $\frac{0.56 A-1.1}{A-3.22}$ & $-$                        & $\frac{0.44A-0.43}{A+0.27}$ \\
$3 s$  & $-$   & $1$   & $-$                       & $-$                         & $\frac{0.45A-0.63}{A-1.4}$ & $-$                         \\
$4 $  & $1$   & $-$   & $\frac{-0.10A-0.01}{A+1}$ & $\frac{0.62A+0.62}{A+1.62}$ & $-$                        & $\frac{0.87A+0.07}{A+2.53}$ \\
    \bottomrule
  \end{tabular}
  \caption{The values of the parameters appearing in DAs of $\Omega_c$ baryon (see Eq.\eqref{eq:17}).}
  \label{tab:1}
\end{table*}
From Eq.\eqref{eq:14}, we obtain the following form using Eq.~\eqref{eq:15}
\begin{equation}
  \label{eq:21}
  \begin{split}
    \Pi_{\mu \nu} &= i \int d^4 x \int_0^{\infty} d\omega \omega \int_0^1 du e^{i(p^\prime - \omega v) x} \bigg\{
                   \bigg[ \sum A_i (Tr \Gamma_i \gamma_\mu) S \mathcal{T}_\nu \gamma_5 \bar{\slashed{v}}
                    - 2 \sum A_i C \Gamma_i^T C^{-1} \gamma_\mu S \gamma_5 \mathcal{T}_\nu \bar{\slashed{v}} \bigg] u_{\Omega_c}~.
 \end{split}
\end{equation}
After integrating over $x$, one can obtain the explicit expressions of the correlation function at QCD level. Separating the coefficients of the Lorentz structures ($ v_\mu \slashed{q} \gamma_\nu \gamma_5 $, $\slashed{q} \gamma_5 v_\mu q_\nu$, $\slashed{q} \gamma_5 v_\mu v_\nu$, $\slashed{q} \gamma_5 g_{\mu \nu}$ ,( $\slashed{q} \gamma_\nu v_\mu$, $\slashed{q} v_\mu q_\nu$, $\slashed{q} v_\mu v_\nu$, $\slashed{q} g_{\mu \nu})$) from both representations of the correlation function,  we get the desired sum rules for the transition form factors $F_1(q^2)$, $F_3(q^2)$, $F_2(q^2) + F_3(q^2)$, and  $F_4(q^2)$ ($(G_1(q^2), G_3(q^2), G_2(q^2) + G_3(q^2))$ and $G_4(q^2)$), respectively.
\begin{align}
  \label{eq:19}
  \frac{\lambda_{\Omega}}{ (p-q)^2 - m_{\Omega}^2 } F_1(q^2) &= \Pi_1~, &    -\frac{\lambda_{\Omega}}{m_{\Omega}^2 - (p-q)^2} G_1(q^2) &= \Pi_5~,  \nonumber \\    
 -\frac{1}{m_\Omega} \frac{\lambda_{\Omega}}{ m_{\Omega}^2 - (p-q)^2 } F_3(q^2) &= \Pi_2~, &  \frac{1}{m_\Omega} \frac{ \lambda_{\Omega}}{m_{\Omega}^2 - (p-q)^2} G_3(q^2) &= \Pi_6~, \nonumber \\
  \frac{\lambda_{\Omega}}{m_{\Omega}^2 - (p-q)^2} \big( F_2(q^2) + \frac{m_{\Omega_c}}{m_\Omega} F_3(q^2) \big) &= \Pi_3~,  &    -\frac{\lambda_{\Omega}}{m_{\Omega}^2 - (p-q)^2} \big( G_2(q^2) + \frac{m_{\Omega_c}}{m_\Omega} G_3(q^2) \big) &= \Pi_7~, \nonumber \\
  \frac{\lambda_{\Omega}}{m_{\Omega}^2 - (p-q)^2} F_4(q^2) &= \Pi_4~, &   -\frac{\lambda_{\Omega}}{m_{\Omega}^2 - (p-q)^2} G_4(q^2) &= \Pi_8~.
\end{align}
Here $\Pi_i$ are the invariant functions for the Lorentz structures mentioned above. In general, the invariant functions can be written in the following form,
\begin{equation}
  \label{eq:22}
  \Pi_i = \int_0^1 du \int d\sigma \sigma \bigg\{ \frac{\rho_i^{(1)}}{\Delta} + \frac{\rho_i^{(2)}}{\Delta^2}  + \frac{\rho_i^{(3)}}{\Delta^3} \bigg\}
\end{equation}
where $\Delta = {p^\prime}^2 - s(\sigma)$, $\sigma =\frac{\omega}{m_{\Omega_c}}$, and $i$ runs from 1 to 8 (each value of the $i$ describes the corresponding structure).
Since the explicit expressions of $\rho_i^{(1)}$, $\rho_i^{(2)}$, and $\rho_i^{(3)}$ are lengthy, we did not present them here.
Applying Borel transformation with respect to the variable $-(p-q)^2$, we get the sum rules for the form factors:
\begin{align}
  \label{eq:23}
  F_{1}                                       & = \frac{e^{m^2_{\Omega}/M^2}}{\lambda_{\Omega}} \Pi^B_{1}~,    & G_{1}                                     & = - \frac{e^{m^2_{\Omega}/M^2}}{\lambda_{\Omega}} \Pi^B_{5}~, \nonumber \\
  F_{2} + \frac{m_{\Omega_c}}{m_\Omega} F_{3}    & = \frac{e^{m^2_{\Omega}/M^2}}{\lambda_{\Omega}} \Pi^B_{3}~,    & G_{2} + \frac{m_{\Omega_c}}{m_\Omega} G_{3} & = -\frac{e^{m^2_{\Omega}/M^2}}{\lambda_{\Omega}} \Pi^B_{7}~, \nonumber   \\
  F_{3}                                       & = - \frac{m_\Omega}{\lambda_\Omega} e^{m_\Omega^2/M^2} \Pi^B_2~, & G_{3}                                       & = \frac{m_\Omega}{\lambda_\Omega} e^{{-m_{\Omega}^2}/M^2} \Pi^B_6~, \nonumber \\
  F_{4}                                       & = \frac{e^{m^2_{\Omega}/M^2}}{\lambda_{\Omega}} \Pi^B_{4}~,    & G_{4}                                       & = -\frac{e^{m^2_{\Omega}/M^2}}{\lambda_{\Omega}} \Pi^B_{8}~.  
\end{align}
Borel transformation and continuum subtraction is performed with the help of formula
\begin{multline}
  \int_0^{\infty} d\sigma \frac{\rho(\sigma)}{ \big[(p-q)^2 - s(u) \big]^n} \Rightarrow 
  \int_0^{\omega_0} d\sigma \bigg\{ (-1)^n \frac{ e^{-s(\sigma)} I_n(u, \sigma)}{(n-1)! (M^2)^{n-1}} \bigg\}                                                                                                           \\
- \frac{(-1)^n e^{-s(\sigma)}/M^2}{(n-1)!} \sum_{j = 1}^{n-1} \frac{1}{(M^2)^{n-j-1}} \frac{1}{s^\prime} (\frac{d}{d\sigma} \frac{1}{s^\prime})^{j-1} I_n |_{\sigma = \sigma_0}~,
\end{multline}
where
\begin{equation}
  \label{eq:24}
  \sigma_0 = \frac{(s_{th} + m_{\Omega_c}^2 - q^2) + \sqrt{(s_{th} + m_{\Omega_c}^2  - q)^2 - 4 m_{\Omega_c}^2 (s_{th} - m_s^2)}}{2 m_{\Omega_c}^2}~,
\end{equation}
and
\begin{equation}
  \label{eq:7}
  I_n = \frac{\rho_n(\sigma)}{\bar{\sigma}^n}~.
\end{equation}
Numerical analysis of the obtained sum rules for the form factors is carried out in the next section.

Having determined the form factors for $\Omega_c^0 \rightarrow \Omega^-$ transition, it is straightforward to calculate the width of the semileptonic $\Omega_c^0 \to \Omega^- l^+ \nu_l$ and non-leptonic $\Omega_c^0 \to \Omega^- \pi^+(\rho^+)$ decays.

First, we present the amplitudes of $\Omega_c^0 \to \Omega^- h$ ($h = \pi, \rho$) and $\Omega_c^0 \to \Omega^- l^+ \nu_l$ in the helicity basis of $H_{\lambda_\Omega \lambda_{h (l)}}$ \cite{Gutsche:2018utw,Zhao:2018mrg,Hsiao:2020gtc}. While $\lambda_\Omega = \pm 3/2, \pm 1/2$ corresponds to the helicity states of the $\Omega$-baryon, $\lambda_{h (l)}$ corresponds to the helicity states of $\pi (\rho)$ and $l^+ \nu_l$ pair, respectively. The helicity amplitudes are defined
\begin{equation}
  \label{eq:25}
  H^{V(A)}_{\lambda_\Omega, \lambda_W} = \langle \Omega(\lambda_\Omega) | \bar{s} \gamma_\mu (\gamma_5) c | \Omega_c(\lambda)  \rangle \epsilon_W^{* \mu} (\lambda_W)~,
\end{equation}
where $\lambda = \lambda_W - \lambda_\Omega$ and $\epsilon_W^{*\mu}$ is the four-vector of the virtual W-boson. Using these definitions of the helicity amplitude(s) and the matrix element  $\langle \Omega | \bar{s} \gamma_\mu (\gamma_5) c | \Omega_c \rangle$ in terms of form the factors, we get the following relations~\cite{Hsiao:2020gtc}.
\begin{equation}
  \label{eq:26}
  \begin{split}
    H^{V(A)}_{3/2,1} &= \mp \sqrt{Q_{\mp}^2} F_4(G_4) \\
    H^{V(A)}_{1/2,1} &= - \sqrt{ \frac{Q_{\mp}^2}{3}}  \big[ F_1(G_1) \frac{Q_{\pm}^2}{m_1 m_2} - F_4(G_4) \big] \\
    H^{V(A)}_{1/2,0} &=   \sqrt{\frac{2 Q_\mp^2}{3 q^2}}  \big[ F_1(G_1) \frac{Q_{\pm}^2 m_{\mp}}{2 m_1 m_2}  \mp \big( F_2(G_2) + F_3(G_3) \frac{m_1}{m_2} \big) \frac{|\vec{p}_\Omega|^2}{m_2} \mp F_4(G_4) \tilde{m}   \big] \\
       \tilde{H}^{V(A)}_{1/2,t} &=   \sqrt{\frac{2 Q^2_{\pm}}{3q^2}} \frac{Q_{\mp}^2 }{2 m_1 m_2}  
     F_1(G_1) m_\pm  \mp \big( F_2(G_2) \tilde{m}_{+} \mp F_3(G_3) \tilde{m}_{-}  \mp F_4(G_4) m_1  \big)
  \end{split}
\end{equation}
In these expressions $m_{\pm} = m_1 \pm m_2$, $Q^2_{\pm} = m^2_{\pm} - q^2$, $\tilde{m}_{\pm} = (m_{+} m_{-} \pm q^2) / 2 m_1 (m_2)$. Here $m_1$ and $m_2$ are the mass of $\Omega_c$ and $\Omega$ baryon, respectively. The remaining helicity amplitudes can be obtained by symmetry relation.
\begin{equation}
  \label{eq:28}
  H^{V(A)}_{-\lambda, -\lambda_W} = \mp H^{V(A)}_{\lambda,\lambda_W}
\end{equation}
From these helicity amplitudes the decay widths of the semileptonic and non-leptonic decays are calculated as
\begin{equation}
  \label{eq:29}
  \Gamma( \Omega_c \to \Omega l \nu) = \frac{G_F^2 |V_{cs}|^2}{192 \pi^3 m^2_{\Omega_c}} \int^{(m_{\Omega_c} - m_\Omega)^2}_{m_l^2} dq^2 \frac{|\vec{p}_\Omega| (q^2 - m_l^2)^2}{q^2} H_l^2~,
\end{equation}
and
\begin{equation}
  \label{eq:30}
  \Gamma( \Omega_c \to \Omega  h) = \frac{G_F^2 |\vec{p}_\Omega | }{32 \pi m^2_{\Omega_c}} |V_{cs} V_{ud}^*|^2 a_1^2 m_h^2 f_h^2 H_h^2
\end{equation}
\begin{align}
  \label{eq:31}
  H_l^2 &= (1 + \frac{m_l^2}{2 q^2}) H_{\rho}^2 +  \frac{3 m_l^2}{2 q ^2} H_\pi^2 
\end{align}
where 
\begin{align}
  \label{eq:31}
  H_\rho^2 &= |H_{3/2,1}|^2 + |H_{1/2,1}|^2 + |H_{1/2,0}|^2 + |H_{-1/2,0}|^2 + |H_{-1/2,-1}|^2 + |H_{-3/2,-1}|^2~, \\
  H_\pi^2 &= |H_{1/2,t}|^2 + |H_{-1/2,t}|^2~. 
\end{align}
Here, $G_F$ is the Fermi coupling constant and $V_{ij}$ are the elements of CKM.
The factor $a_1 = C_1 + C_2 / N_c$ comes from the factorization~\cite{Hsiao:2019ann}, where $N_c$ is the color factor and $C_1 = -0.25 $ and $C_2= 1.1$ are the Wilson coefficients~\cite{Gutsche:2018utw}.
\section{Numerical Analysis}
\label{sec:3}
In this section, we perform numerical analysis for the transition form factors obtained in the previous section. For this goal, first, we present the input parameters in our numerical analysis.
\begin{align}
  \label{eq:13}
  m_c({1 GeV}) & = 1.35 \pm 0.10~{GeV}~\text{\cite{ParticleDataGroup:2020ssz}} & m_s (1~{GeV}) & = 0.12 \pm 0.02~{GeV}~\text{\cite{ParticleDataGroup:2020ssz}} \nonumber \\
  m_{\pi}         & = 0.140~{GeV}~\text{\cite{ParticleDataGroup:2020ssz}}         & m_{\rho}         & = 0.77~{GeV}~\text{\cite{ParticleDataGroup:2020ssz}} \nonumber  \\
  m_{\Omega_c}    & = 2.695~{GeV}~\text{\cite{ParticleDataGroup:2020ssz}}         & m_\Omega         & = 1.672~{GeV}~\text{\cite{ParticleDataGroup:2020ssz}} \nonumber \\ 
  f_\pi           & = 132~{MeV}~\text{\cite{ParticleDataGroup:2020ssz}}           & f_\rho           & = 216~{MeV}~\text{\cite{ParticleDataGroup:2020ssz}} \nonumber \\
  f^{(1)}         & = 0.093 \pm 0.01~\text{\cite{Wang:2009cr}  }                     & f^{(2)}          & = 0.093 \pm 0.01~\text{\cite{Wang:2009cr}} \\
  V_{cs}          & = 0.97320 \pm 0.00011~\text{\cite{ParticleDataGroup:2020ssz}}    & V_{ud}           & = 0.97401 \pm 0.00011~\text{\cite{ParticleDataGroup:2020ssz}}
\end{align}
For the quark masses, $\overline{MS}$ scheme values are used.
Besides these input parameters, LCSR involve two auxiliary parameters, the continuum threshold $s_{th}$ and Borel mass parameter $M^2$. The working region of $M^2$ is determined in a way that the power corrections as well as the continuum contributions are suppressed. The working region of $s_{th}$ is determined from the condition that the mass sum rules reproduce the mass with say $10\%$ accuracy. Following these criteria, we obtain the working regions of $s_{th}$ and $M^2$:
\begin{align}
  \label{eq:32}
  &4.0~\rm{GeV^2} \leq s_{th} \leq 4.5~\rm{GeV^2} \nonumber \\
  &3.0~\rm{GeV^2} \leq M^2 \leq 4.0~\rm{GeV^2} 
\end{align}
In these working regions of $s_{th}$ and $M^2$, both the conditions of the smallness of the sub-leading twist-3, twist-4 contributions and the suppression of the higher states as well as continuum contributions are satisfied.

Having determined the working intervals of the threshold and Borel mass parameters, the next problem is to find the best fitting for the form factors. The LCSR predictions for the form factors are not applicable for the whole physical region, $m_l^2 \leq q^2 \leq (m_{\Omega_c}^2 - m_\Omega)^2$ but give reliable results for up to $q^2 \leq 0.5~\rm{GeV^2}$ region. Hence, we first obtained the form factors within QCD sum rules up to $q^2 \simeq 0$. Then, we  extrapolated from the domain where LCSR predictions are reliable to the full physical region by applying the following z-expansion fit function~\cite{Bharucha:2015bzk,Bourrely:2008za}. 
\begin{equation}
  \label{eq:10}
  F_i(q^2) = \frac{1}{1 - \frac{q^2}{m_{R,i}^2}} \bigg( a_0^{i} + a_1^{i} (z(q^2) - z(0)) + a_2^{i} (z(q^2) - z(0))^2 \bigg) ~.
\end{equation}
where 
\begin{equation}
  \label{eq:9}
  z(t) = \frac{\sqrt{t_{+} - t} - \sqrt{t_+ - t_0}}{\sqrt{t_{+} - t} + \sqrt{t_+ - t_0}}
\end{equation}
and $t_\pm = (m_{\Omega_c} \pm m_{\Omega})^2$,  $t_0 = t_{+} (1- \sqrt{1- \frac{t_{-}}{t_{+}}} )$.

Here, $m_{R,i}$ are the corresponding masses of the resonances for the $c \to s$ transition in the spectrum, i.e., $m_{D_s} = 1.97~\rm{GeV}$. 

The obtained parametrization that best reproduces the form factors predicted by the LCSR in the region $q^{2} \leq 1.1 ~\rm{GeV}^{2}$, is given in Table~\ref{tab:2}. Note that $a_0$ corresponds to the form factor at $q^2=0$, i.e., $a_0 = F_i(q^2=0)$.
\begin{table*}[t]
  \centering
  \renewcommand{\arraystretch}{1.4}
  \setlength{\tabcolsep}{3.2pt}
  \begin{tabular}{lccc}
    \toprule
                 & $a_0 $ & $a_1$ & $a_2$  \\ 
    \midrule
        $F_1$ & $-0.55 \pm 0.05 $ & $6.391$ & $-191.2$ \\
        $F_2$ & $-0.68 \pm 0.07 $ & $30.94$ & $-1053$ \\
        $F_3$ & $1.0 \pm 0.2 $ & $-35.60$ & $1117$ \\
        $F_4$ & $0.16 \pm 0.02 $ & $-7.24$ & $239.9$ \\
        $G_1$ & $-0.48 \pm 0.02 $ & $3.513$ & $-100.9$ \\
        $G_2$ & $0.68 \pm 0.07 $ & $-30.1$ & $1053$ \\
        $G_3$ & $-1.0 \pm 0.2 $ & $35.60$ & $-1117$ \\
        $G_4$ & $-0.16 \pm 0.02 $ & $7.24$ & $-239.9$ \\
    \bottomrule
  \end{tabular}
  \caption{The values of the form factors at $q^2 =0$ and the fit parameters of $a_i$.}
  \label{tab:2}
\end{table*}
%
%
\begin{figure}
\includegraphics[width=0.49\textwidth]{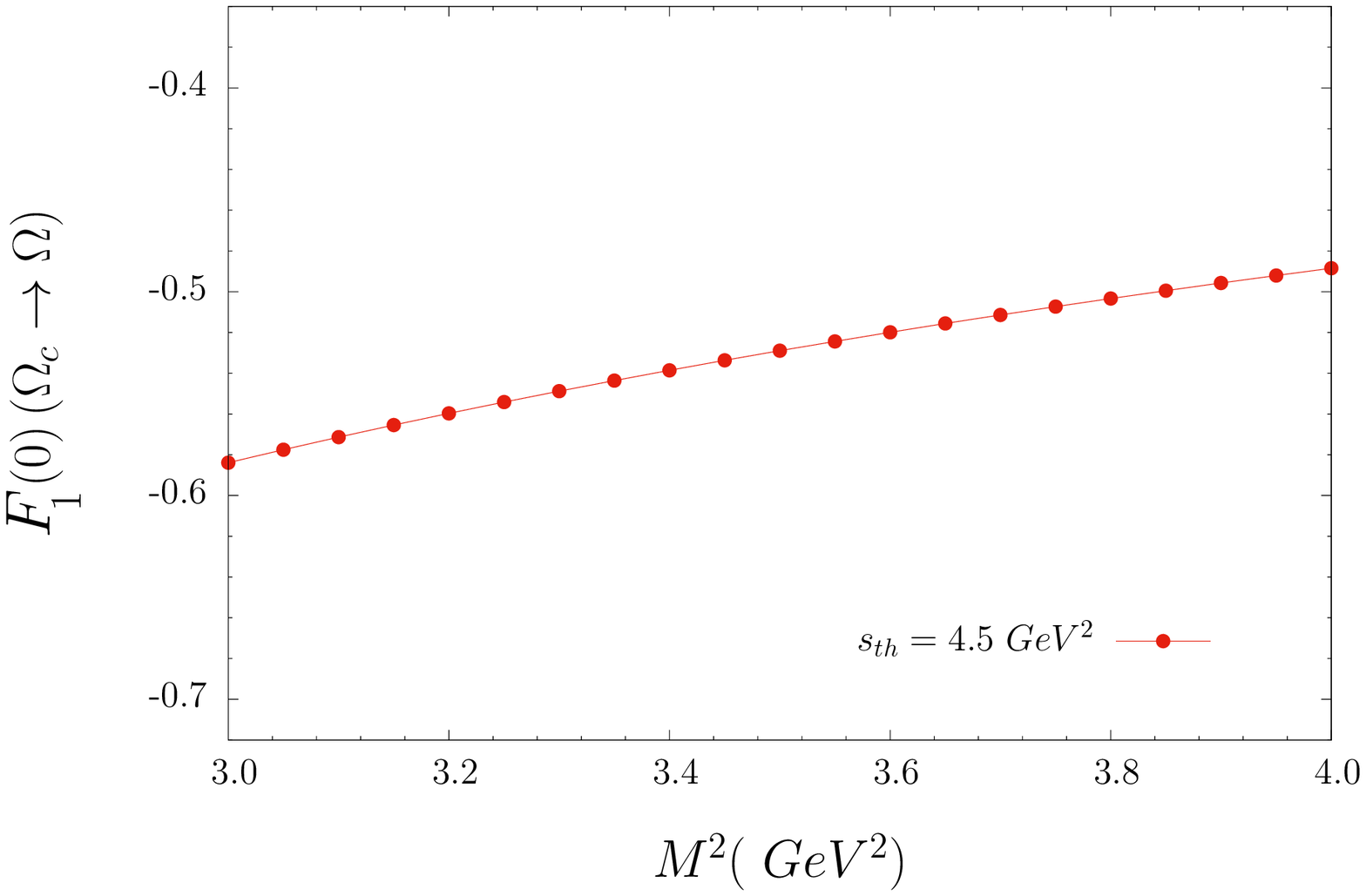} 
\includegraphics[width=0.49\textwidth]{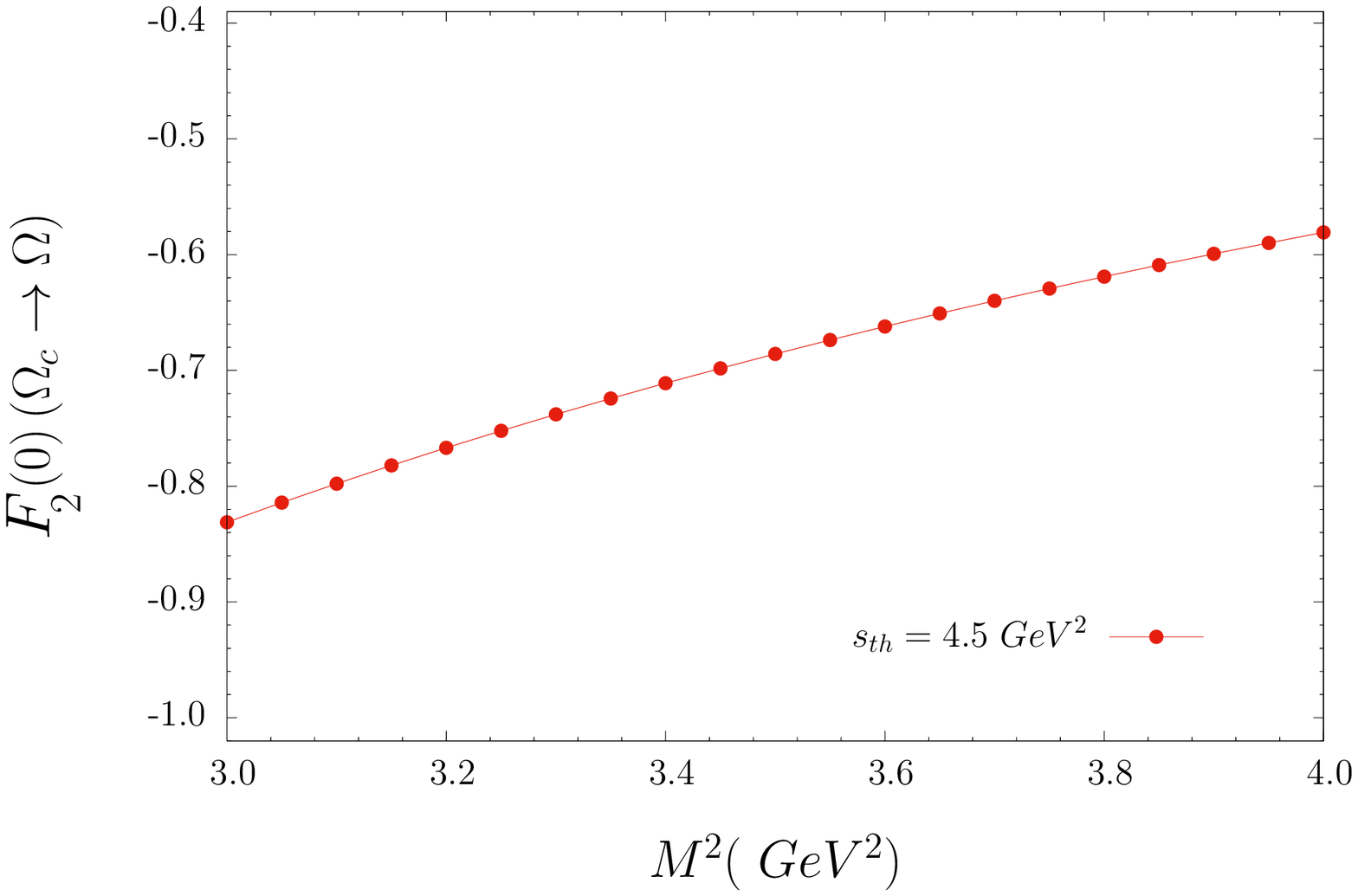} \\
\includegraphics[width=0.49\textwidth]{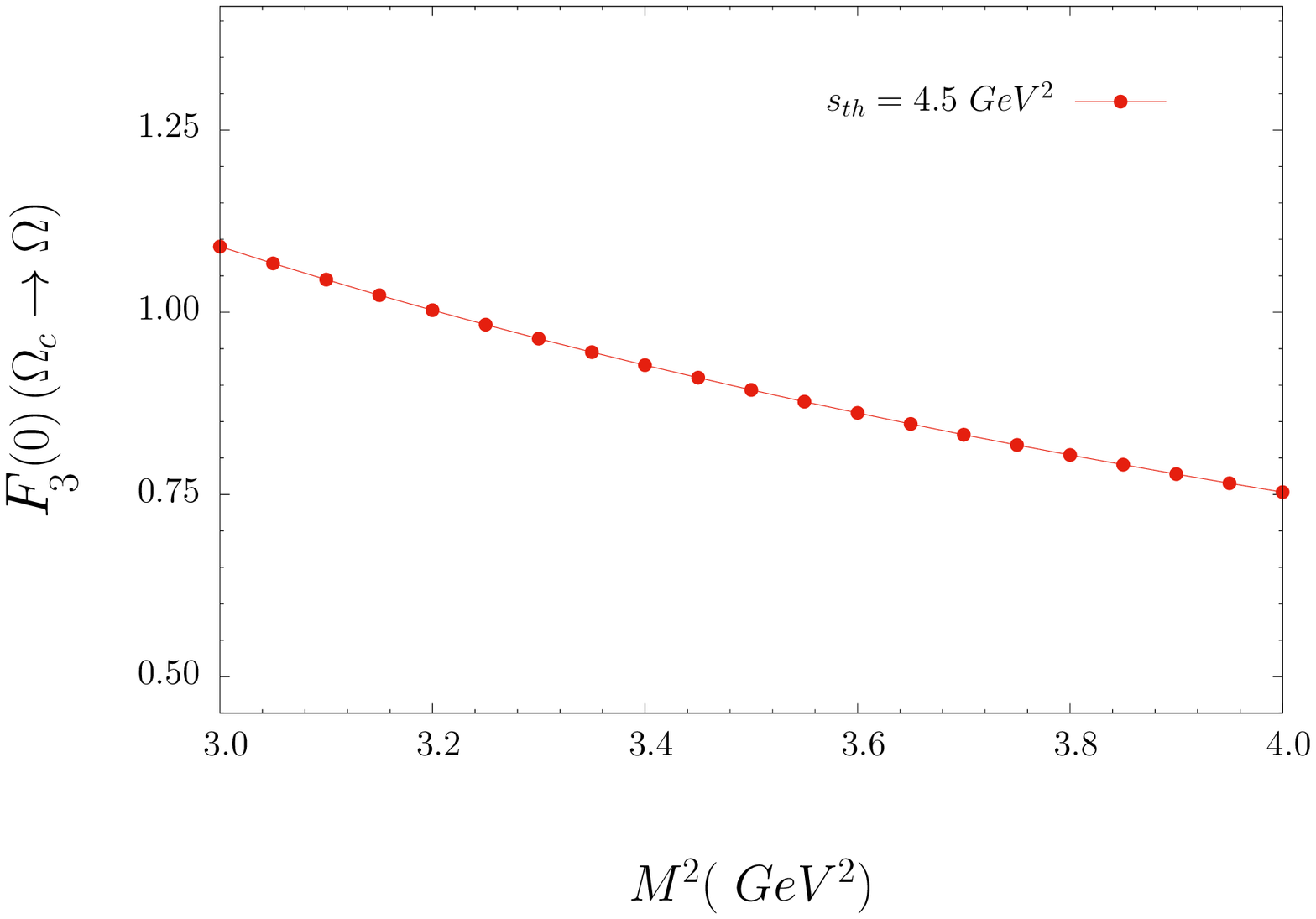}
\includegraphics[width=0.49\textwidth]{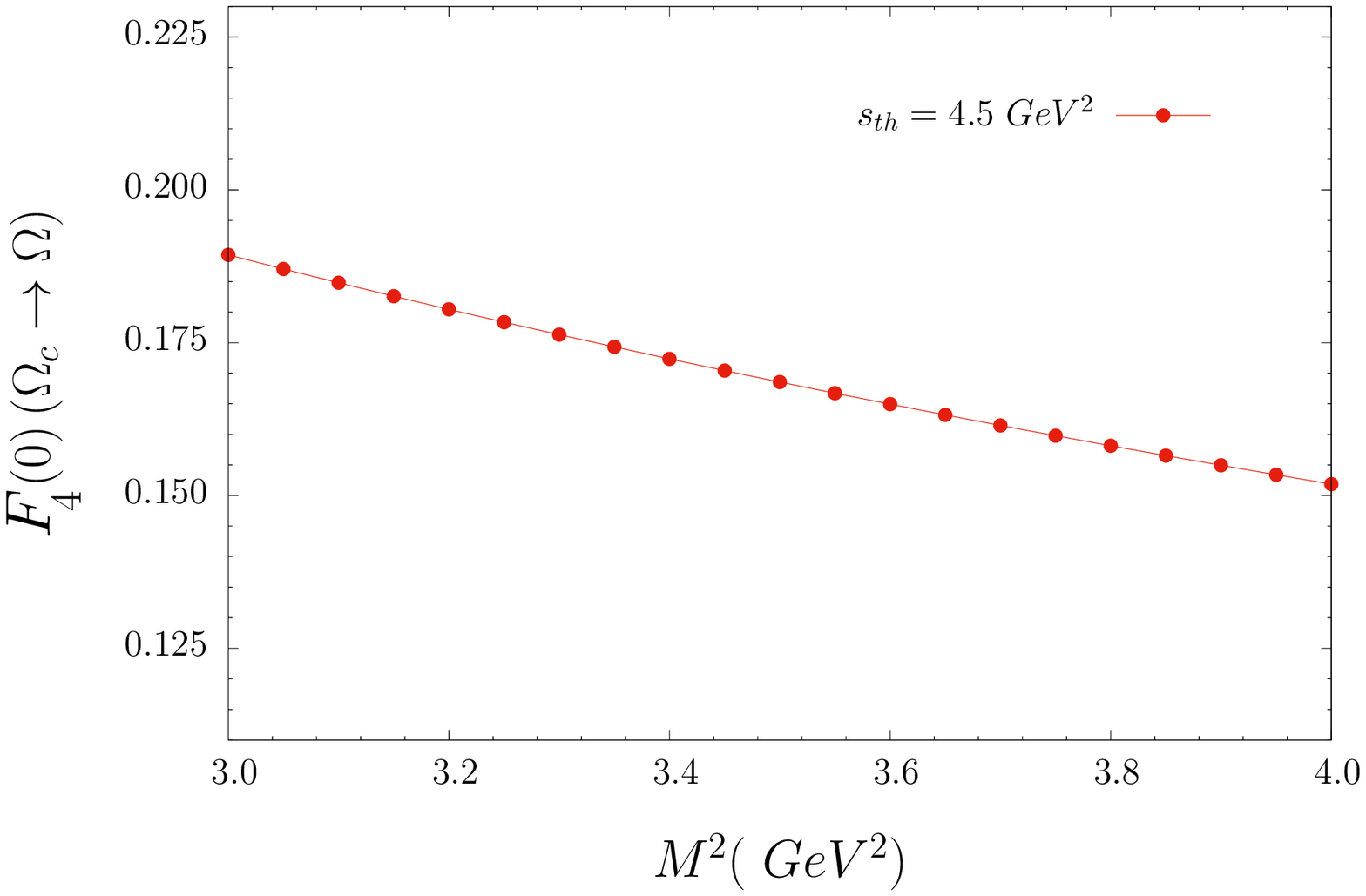}\\
\includegraphics[width=0.49\textwidth]{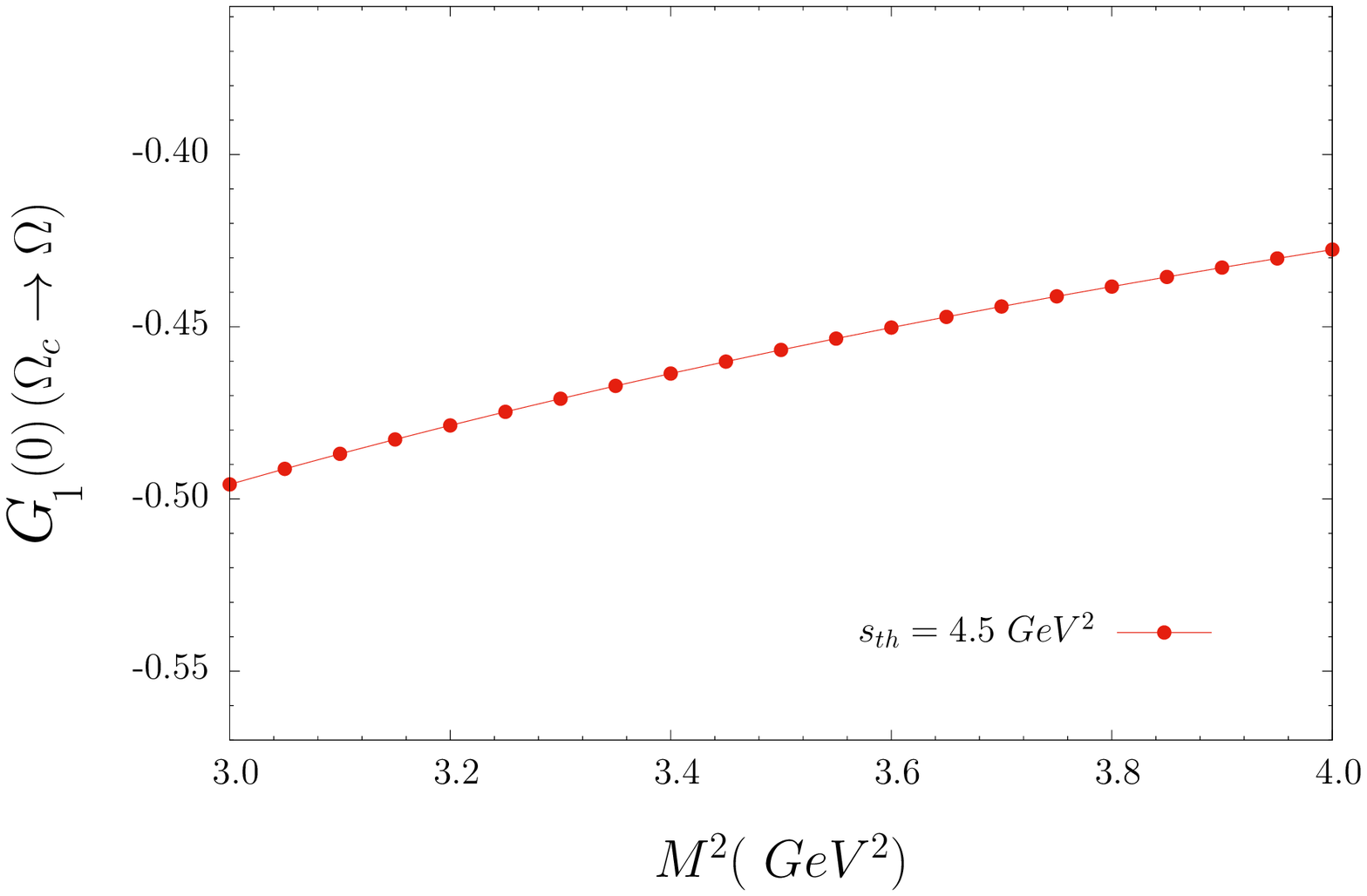} 
\includegraphics[width=0.49\textwidth]{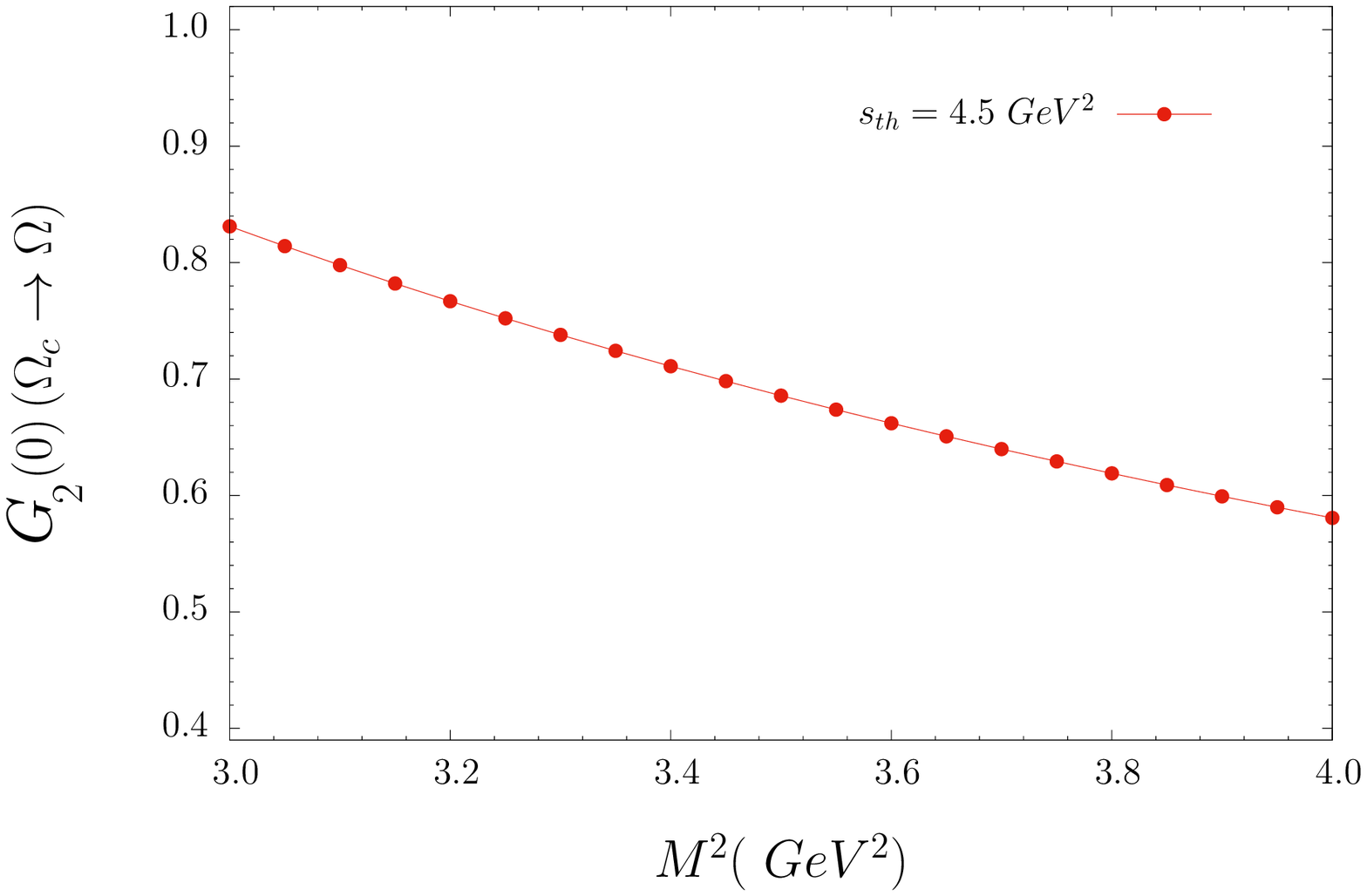} \\
\includegraphics[width=0.49\textwidth]{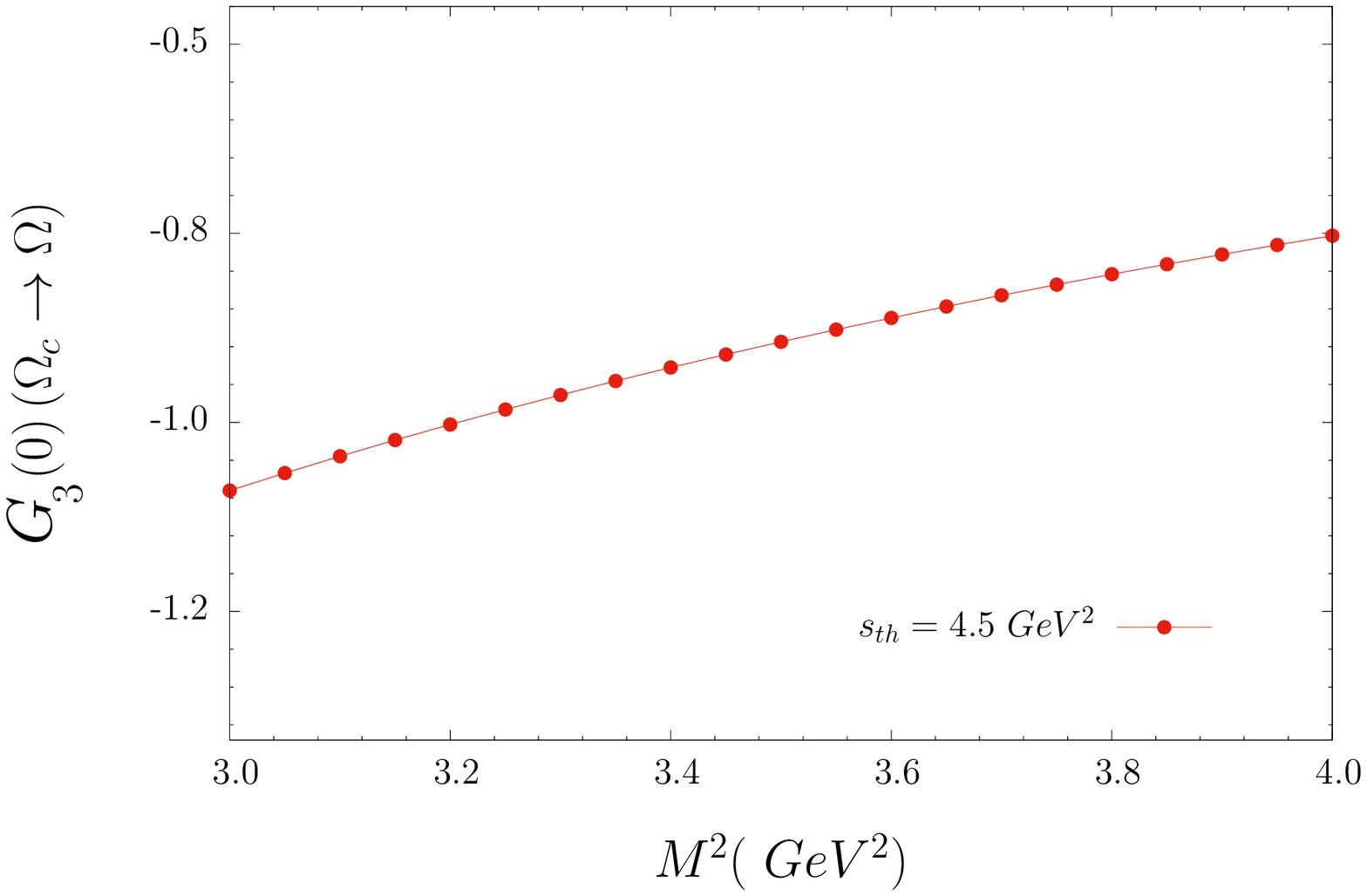}
\includegraphics[width=0.49\textwidth]{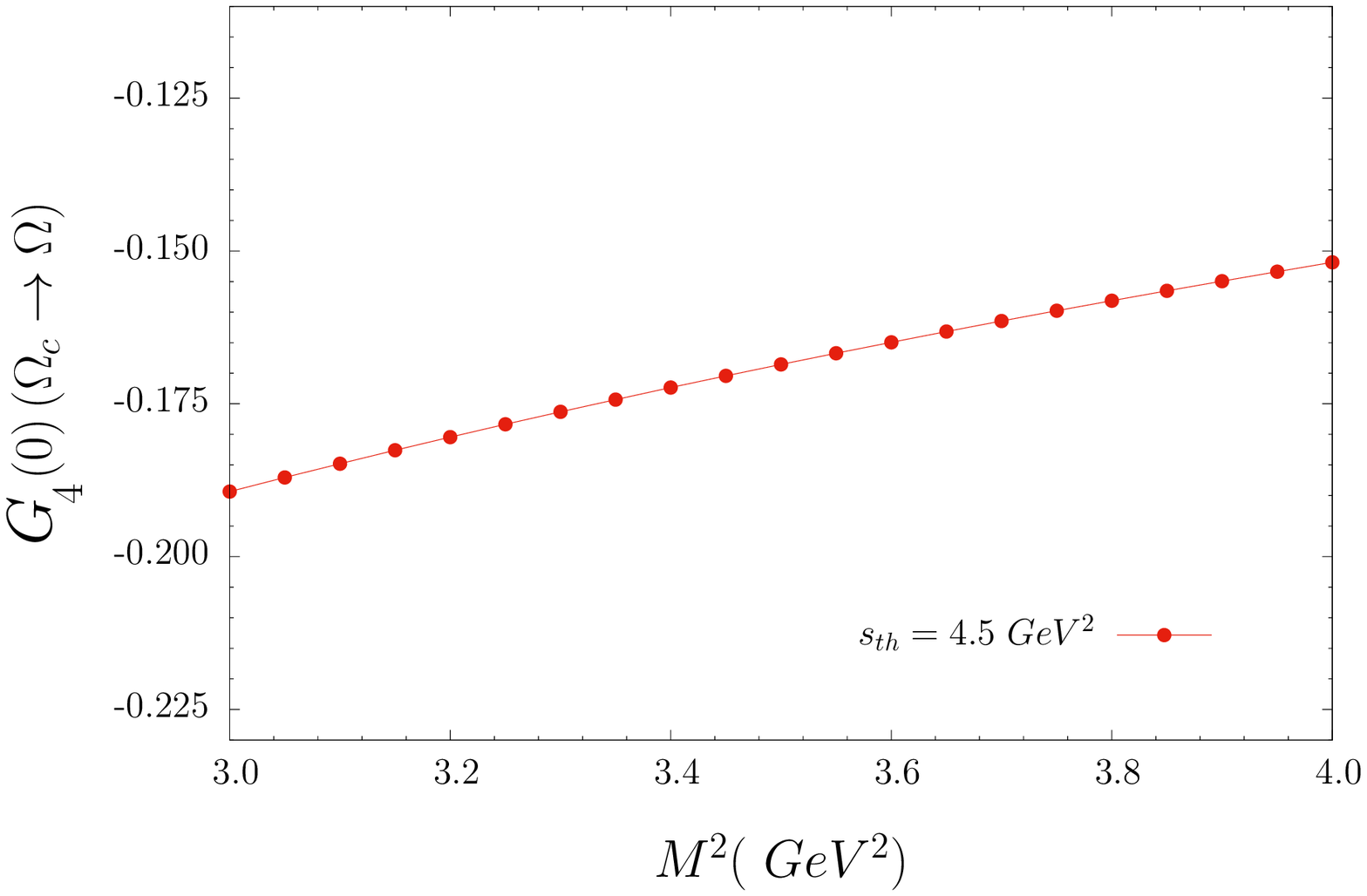}
\caption{The working regions of $M^2$ and $s_0$ as well as the value of residues for the considered states.}
\label{fig1}
\end{figure}
%
To verify that the results of the form factors depend weakly on the chosen $M^2$ and $s_{th}$ auxiliary parameters, we plotted the variation of the form factors at $q^2=0$, on $M^2$ and $s_{th} = 4~\rm{GeV^2}$ in Figure~\ref{fig1}. The figure shows good stability of the form factors on $M^2$ and $s_{th}$.

Using the obtained results for the form factors responsible for $\Omega_c^0 \to \Omega^-$ transition, we can calculate the branching ratio for $\Omega_c^0 \to \Omega^- l^+ \nu_l$ and $\Omega_c^0 \to \Omega^- h^+$ ($h=\pi,\rho$) decay by using Eqs.\eqref{eq:9} and \eqref{eq:10}. The lifetime of the $\Omega_c$ baryon is taken as $\tau = (268 \pm 24 \pm 10) \times 10^{-15}~\rm{s}$ \cite{ParticleDataGroup:2020ssz}. The Belle II collaboration has recently reported a lifetime of $\Omega_c$ baryon, $\tau = 243 \pm 48 \pm 11$, which agrees with the previous measurements~\cite{Belle-II:2022plj}. Using the values of the input parameters together with the decay width expressions, we obtain the branching ratios that are presented in Table~\ref{tab:son}.

$\Omega_c^0 \to \Omega^-$ transition had been studied in several models~\cite{Hsiao:2020gtc,Xu:1992sw,Cheng:1996cs,Gutsche:2018utw,Pervin:2006ie}. The obtained results are presented in Table~\ref{tab:son}. Our results are close to the predictions of the light-front quark model~\cite{Hsiao:2020gtc} for the semileptonic part. However, there is a large discrepancy with other studies for these decays. In addition, our theoretical predictions do not match with the recent BELLE II measurement~\cite{Belle:2021dgc}, especially for the semileptonic $\Omega_c^0 \to \Omega^- l^+ \nu_l$ decay. A similar situation was also obtained in~\cite{Hsiao:2020gtc}. This discrepancy needs further discussion, and it may indicate the existence of new physics. On the other hand, an upper bound for $\mathcal{R_{\rho/\pi}} > 1.3$ is set by the experiment. This bound is not controversial with theoretical results; however, there is considerable discrepancy among the predictions of different theoretical approaches. Hopefully, more experimental and theoretical efforts on this transition will enlighten the discrepancy.
\begin{table*}[t]
  \small
  \centering
  \renewcommand{\arraystretch}{1.2}
  \setlength{\tabcolsep}{3pt}
  \begin{tabular}{lccccccc}
    \toprule
    States                     & Our Result & Exp.~\cite{Belle:2021dgc,ParticleDataGroup:2020ssz} & Ref~\cite{Hsiao:2020gtc}         & Ref~\cite{Xu:1992sw}   & Ref~\cite{Cheng:1996cs} & Ref~\cite{Gutsche:2018utw} & ~\cite{Pervin:2006ie} \\ 
    \midrule
    $\mathcal{B}_\pi$          & $29 \times 10^{-3}$         & ...         & $(6 \pm 0.8) \times 10^{-3}$     & $66.5 \times 10^{-3}$  & $42.3 \times 10^{-3}$     & ...                        &                       \\
    $\mathcal{B}_\rho$         & $63 \times 10^{-3}$         & ...        & $(17 \pm 0.5) \times 10^{-3}$    & $361.1 \times 10^{-3}$ & $149 \times 10^{-3}$      & ...                        &                       \\
    \midrule
    $\mathcal{B}_e$            & $20.6 \times 10^{-3}$          & ...         & $(5 .4 \pm 0.2) \times 10^{-3} $ & ...                    & ...                     & ...                        & $127 \times 10^{-3}$  \\ 
    $\mathcal{B}_\mu$          & $19.6 \times 10^{-3}$         & ...         & $(5 .0 \pm 0.2) \times 10^{-3} $ & ...                    & ...                     & ...                        &                       \\
    \midrule
    $\mathcal{R}_{\rho / \pi}$  & $2.18$         & $>1.3$         & $( 2.8 \pm 0.4)$                 & $5.4$                  & $3.5$                   & $9.5$                      &                       \\ 
    $\mathcal{R}_{e / \pi}$    & $0.71 $         & $1.98 \pm 0.13 \pm 0.08$          & $(0.9 \pm 0.1)$                  & ...                    & ...                     & ...                        &                       \\
    $\mathcal{R}_{\mu / \pi}$    & $0.68$         & $1.94 \pm 0.18 \pm 0.10$          & $(0.9 \pm 0.1)$                  & ...                    & ...                     & ...                        &                       \\
    \bottomrule
  \end{tabular}
  \caption{Branching fractions of the $\Omega_c^0$ decays obtained via different models as well as experimental results are presented. $\mathcal{R}$ corresponds to the branching ratio of the considered decays.}
  \label{tab:son}
\end{table*}
\section{Conclusion\label{conclusion}}
In this work, the semileptonic and non-leptonic decays of charged decays of $\Omega_c^0$ , namely, $\Omega_c^0 \rightarrow \Omega^- l^+ \nu_l$,  $\Omega_c^0 \rightarrow \Omega^- \pi^+$ and $\Omega_c^0 \rightarrow \Omega^- \rho^+$  are studied within the framework of the light-cone sum rules by using the distribution amplitudes of $\Omega_c$ baryon. In this study, DAs for $\Omega_b$ baryons are used for their c-quark counterpart depending on the heavy-quark symmetry. We first calculated the transition form factors for $\Omega_c^0 \to \Omega^- $ decay in the LCSR method. Then, using the obtained results for the transition form factors, we predicted the branching ratios of the semileptonic $\Omega_c^0 \rightarrow \Omega^- l^+ \nu_l$ (where $l = e, \mu$) and non-leptonic  $\Omega_c^0 \rightarrow \Omega^- \pi^+ (\rho^+)$ decays. Finally, we compared our predictions with other approaches as well as recent BELLE II results. We obtained that our results, especially on branching ratio for semileptonic $\Omega_c^0 \rightarrow \Omega^- l^+ \nu_l$ decay normalized to $\Omega_c^0 \to \Omega^- \pi^+$ is considerably smaller than existing experimental data. A similar discrepancy was also obtained in the light-front approach.

We obtained that there is a large deviation between the theoretical predictions and experimental results on the branching ratios of the semileptonic decays. Our results on branching ratios for the semileptonic decay (normalized to $\Omega \to \pi$) is compatible with the light-front quark model. However,  other approaches' results drastically differ from ours. Similar circumstances take place for the non-leptonic decays too. Although there is only a lower bound for the ratio, $\mathcal{R}_{\rho / \pi}$, theoretical predictions differ among themselves. 
At this stage, it is hard to identify the reason for the discrepancies for both semileptonic and non-leptonic decays of $\Omega_c$ baryon. These points need further studies from both experimental and theoretical sides, and even new physics implications may be implied.

\bibliographystyle{utcaps_mod}
\bibliography{../../all.bib}


\end{document}